\documentclass[prb,twocolumn,showpacs,amsmath,amssymb,floatfix,superscriptaddress]{revtex4}
\usepackage[toc,page]{appendix}
\usepackage{graphicx,pdflscape,epsf,epsfig,amsmath}
\usepackage{alltt,dsfont,amsmath,amssymb,bm}
\usepackage{float}

\usepackage{hyperref}
\usepackage{color}

\newcommand{\iden}{ \mathds{ 1}}

\newcommand{\A}{{\cal A}}

\newcommand{\G}{{\cal{G}}}

\newcommand{\GH}{{\bf g}}
\newcommand{\GHI}{\GH^{-1}}

\newcommand{\X}[2]{X_{{#1}}^{#2}}
\newcommand{\si}{\sigma}
\newcommand{\sib}{\bar{\sigma}}

\newcommand{\tJ}{\ $t$-$J$ \ }

\newcommand{\U}{{\cal U}}

\newcommand{\V}{{\cal V}}

\newcommand{\bb}[1]{{{\mathbf #1}}}

\newcommand{\nn}{\nonumber}
\newcommand{\chem}{{\bm \mu}}

\newcommand{\barray}{\begin{eqnarray}}
\newcommand{\earray}{\end{eqnarray}}
\newcommand{\beq}{\begin{eqnarray}}
\newcommand{\eeq}{\end{eqnarray}}

\newcommand{\disp}[1]{Eq.~(\ref{#1})}
\newcommand{\refdisp}[1]{Ref.~(\onlinecite{#1})}

\newcommand{\vk}{\vec{k}}

\newcommand{\lab}[1]{\label{#1}}

\begin{document}
\title{Extremely Correlated Fermi Liquids in the limit of infinite dimensions}
\author{Edward Perepelitsky and B. Sriram Shastry}
\affiliation{ Physics Department, University of California, Santa Cruz, CA 95064, USA}
\date{\today}
\begin{abstract}
We study  the infinite spatial dimensionality limit  ($d \to \infty$) of the
 recently developed Extremely Correlated Fermi Liquid (ECFL) theory\cite{ECFL,Monster} for the \tJ model at $J=0$.  We directly analyze the Schwinger equations of motion for the  Gutzwiller projected  (i.e. $U=\infty$) electron Green's function
$\G$. From simplifications arising in this limit $d \to \infty$,
we are able to make several exact statements about the theory.  The ECFL Green's function  is  shown to have a   momentum independent  Dyson (Mori) self energy. For practical calculations we introduce a partial projection parameter $\lambda$, and obtain the complete set of  ECFL integral equations to $O(\lambda^2)$. In a related publication \cite{ECFLDMFT}, these equations are compared in detail with the dynamical mean field theory for  the large $U$ Hubbard model. Paralleling the well known mapping for the Hubbard model, we find that 
the infinite dimensional \tJ model (with $J=0$)  can be mapped to the infinite-U Anderson impurity model with a self-consistently determined set of parameters. This mapping extends individually to the auxiliary Green's function $\GH$ and the caparison factor $\mu$. Additionally, the optical conductivity is shown to obtainable from $\G$ with negligibly small vertex corrections. These results are shown to hold to each order in $\lambda$.



   \end{abstract}
\maketitle


\section{Introduction\lab{sec1} }

\subsection{ Motivation \label{model}}
The Hubbard model (HM)  with the Hamiltonian:
\beq H = -\sum_{ij\sigma}t_{ij}c^\dagger_{i\sigma }c_{j\sigma} +U\sum_{i}n_{i\uparrow}n_{i\downarrow}- \chem\sum_{i}n_i, \eeq
has attracted  great theoretical interest in condensed matter physics, and is also a fairly realistic  model of strongly correlated materials such as the cuprates. While the small $\frac{U}{t}$ limit  is well described by standard Fermi-Liquid theory\cite{Landau1,Landau2}, the large and intermediate $\frac{U}{t}$ (strongly correlated) cases are much less well understood. Considerable progress has been made by considering the HM in the limit of infinite dimensions \cite{Kuramoto, MetznerVollhardt, Muller-Hartmann,Metzner,Ohkawa,GeorgesKotliar,GeorgesKotliaretal,Khurana,ZlaticHorvatic}. One important result  is that the Dyson self energy, defined by inverting  the expression for the electron Green's function $\G$: 
\beq \G(k) = \frac{1}{i\omega_k + \chem-\epsilon_k-\Sigma_D(k)}\lab{GrD}, \eeq 
becomes momentum independent in this limit \cite{Kuramoto,MetznerVollhardt,Muller-Hartmann,Metzner}. Two other important results are the self-consistent mapping of the infinite dimensional HM  onto the Anderson Impurity model (AIM), detailed in \cite{GeorgesKotliar}(Dynamical Mean Field Theory), and the vanishing of the vertex corrections in the optical conductivity\cite{Khurana,ZlaticHorvatic}, so that the two particle response is obtainable from the single particle Green's function. The  Dynamical Mean Field Theory (DMFT) provides a means for doing reliable numerical calculations for the Hubbard model, at any value of $U$ and has continued to provide new, and interesting results\cite{Xengetal,OtsukiVollhardt}.

A different approach to understanding strong correlations is to consider the extreme correlation limit, where on sets $U \to \infty$ at the outset. In this case, the Hilbert space is Gutzwiller projected so that only single occupancy is allowed on each lattice site. One such extremely correlated model, the \tJ model, consists of taking the $U\to \infty$ limit of the Hubbard model (the t part of the model) and adding on a nearest neighbor anti-ferromagnetic coupling term (the J part of the model).  The $t$ model studied here, is obtained by dropping the $J$ term and thus is identical to the $U= \infty$ limit of the HM. It has been argued by Anderson\cite{Anderson} that the \tJ model describes the physics of the cuprates,  thereby providing an impetus for its detailed study. The Hamiltonian for this model can be written in terms of the Hubbard $X$ operators as\cite{ECQL}
\barray
 H &=& -\sum_{ij\sigma}t_{ij}X_i^{\sigma 0 }X_j^{0\sigma} - \chem\sum_{i\sigma}X_i^{\sigma\sigma} +\frac{1}{2}\sum_{ij\sigma}J_{ij}X_i^{\sigma\sigma} \nn\\ 
 && + \frac{1}{4}\sum_{ij\sigma_1\sigma_2}J_{ij}\{X_i^{\sigma_1\sigma_2 }X_j^{\sigma_2\sigma_1} -X_i^{\sigma_1\sigma_1 }X_j^{\sigma_2\sigma_2}\}.\label{tJmodel}
\earray
The operator $X_i^{ab} = |a\rangle\langle b|$ takes the electron at  site $i$ from the state $|b\rangle$ to the state $|a\rangle$, where $|a\rangle$ and $|b\rangle$ are one of the three allowed states $|\uparrow\rangle$, $|\downarrow\rangle$, or $|-\rangle$.
Our present goal  is to obtain a formally exact solution of the  above $t$ model in the limit of large dimensions by studying its equations of motion. This is designed to be methodologically independent of the available  DMFT solution of the HM  with $U= \infty$, and can be compared with it. 

Our object of study is  the Green's function, written as
\beq \G_{\sigma_1\sigma_2}(i,f) = -\langle T_\tau X_{i}^{0\sigma_1}(\tau_i) X_{f}^{\sigma_2 0}(\tau_f)\rangle, \eeq
where the angular brackets indicate the usual thermal average.
Due to the non-canonical commutation relations of the $X$ operators, the high frequency limit of the Green's function is $\frac{1-\frac{n}{2}}{i\omega_n}$ rather than $\frac{1}{i\omega_n}$ as in the canonical case. To avoid linear growth of the self-energy in the high frequency limit\cite{ECQL}, the Dyson self-energy must be redefined to the Dyson-Mori self energy \cite{Anatomy} as in:
\beq \G(k) = \frac{1-\frac{n}{2}}{i\omega_k + \chem-\epsilon_k(1-\frac{n}{2})-\Sigma_{DM}(k)} \lab{GrDys}.\eeq
Just as is the case for $\Sigma_D$ in the finite-U Hubbard model, $\Sigma_{DM}$ is finite as $i\omega \to \infty$ in the \tJ model.

Shastry has recently introduced a novel and promising approach for calculating correlation functions within the \tJ model based on Schwinger's formulation of field theory. \cite{ECQL,ECFL,Monster} This has culminated in the theory of the Extremely Correlated Fermi Liquid (ECFL) \cite{ECFL,Monster}. This theory has been successfully benchmarked against: line shapes from (ARPES) experiments\cite{Gweon,Kazue}, high-temperature series\cite{Moments} and the numerical renormalization group  (NRG) calculations for the Anderson impurity model\cite{ECFLAM}. A recent  theoretical benchmarking related to this work is the comparison with  DMFT calculations for the large $U$ Hubbard model in a concurrent publication\cite{ECFLDMFT}, with the formulas found here. Indeed the main motivation of the present paper is to obtain results in the limit of large $d$ for the same model, the \tJ model (at $J=0$) or equivalently the  $U= \infty $  Hubbard model 
by  two different methods,   the ECFL  and the DMFT, allowing   such a comparison.

In the ECFL theory, the physical green's function $\G(k)$ is factored into a canonical auxiliary Green's function $\GH(k)$ and an adaptive spectral $\mu(k)$, where $k=(\vec{k}, i \omega_k)$. 
\beq \G(k) = \GH(k) \times \mu(k). \label{Ggmu}\eeq 
These two factors are in turn written in terms of two self-energies, $\Phi(k)$ and $\Psi(k)$.
\beq \GHI(k)= i\omega_k + \mbox{\boldmath{$\mu$}} - (1-n/2)\epsilon_k -\Phi(k), \label{gink} \eeq
\beq \mu(k) = 1-\frac{n}{2} + \Psi(k). \label{muk}\eeq
Here $\Phi(k)$ plays the role of a Dyson self-energy for the canonical Green's function $\GH(k)$, and $\Psi(k)$ is a frequency-dependent correction to $\mu(k)$ from its high frequency value of $1-\frac{n}{2}$. $\Phi$ and $\Psi$ are then given in terms of the vertices
( i.e. functional derivatives w.r.t. the source of the   $\GHI$ and $\mu$) as will be described below, leading to a closed set of Schwinger differential equations (the ECFL equations of motion). These equations are in general intractable since there is no obvious  small parameter, 
 and therefore to enable practical calculations, an expansion is carried out in a partial projection parameter $\lambda$.  Here $\lambda$   interpolates between the free Fermi gas and the \tJ model. The meaning of $\lambda$ as a partial projection parameter is detailed  in \cite{Monster}, and may be summarized in the mapping $\X{i}{\si 0} \to f_{i \si}^\dagger ( 1- \lambda \ n_{i \sib})$, where $f_{i \si}$ is a canonical electron operator. Thus at $\lambda=0$ we have canonical electrons, whereas at $\lambda=1$ we have the fully projected electrons.

In this work, our aim is to combine the two approaches, namely to consider the ECFL in the limit of infinite spatial dimensions. In this limit, $J\to0$, and the infinite-dimensional \tJ model becomes the infinite dimensional infinite U Hubbard model (see sec. 6A of \refdisp{ECFLDMFT} for a brief discussion of this). It is not clear a priori, whether or not the aforementioned results, valid for the infinite dimensional finite-U Hubbard model, carry over to the infinite dimensional \tJ model. The possible conflict arises from the fact that in the case of the former, the ratio $\frac{U}{d}\to0$, while in the case of the latter, $\frac{U}{d}\to\infty$. This question was raised in \refdisp{Response}, pointing to the ECFL solution of the infinite dimensional \tJ model as a source of resolution.
 Working directly with the infinite-U Hamiltonian (\disp{tJmodel} with $J=0$), and using the corresponding ECFL equations of motion, we are able to address this challenging task and to 
 show that the two limits $U\to\infty$ and $d\to\infty$ do in fact commute.

Moreover, we are able to determine the structure of the ECFL objects $\Phi(k)$ and $\Psi(k)$ in the limit of infinite dimensions. Such structural information has already been used to fit numerical results obtained through DMFT calculations to a convenient and flexible functional form\cite{ECFLDMFT}. Finally, we are able to elucidate the nature of the $\lambda$ expansion in the large d limit. For readers who might be  more interested in the results than the methodology,
we  provide a detailed summary of our results at the outset.

\subsection{Results in the limit of infinite dimensions\label{ECFLGr}} 

We show that in the large d limit, the two self energies $\Phi(k)$ and $\Psi(k)$ simplify in the following way.
\beq \Psi(k) = \Psi(i\omega_k), \lab{Psiid} \\
 \Phi(k) = \chi(i\omega_k)+\epsilon_k\Psi(i\omega_k). \lab{Phiid}\eeq
These in turn shows that the Dyson-Mori self energy behaves as
\beq \Sigma_{DM}(k)=\Sigma_{DM}(i\omega_k)= \frac{(i\omega_k + \mbox{\boldmath{$\mu$}})\Psi(i\omega_k)+(1-\frac{n}{2})\chi(i\omega_k)}{1-\frac{n}{2}+\Psi(i\omega_k)}, \nn \\
\label{selfrelatemin2} 
 \eeq
and is therefore local in the limit of infinite dimensions. We show that to each order in the $\lambda$ expansion, $ \Psi(i\omega_k)$ and $ \chi(i\omega_k)$ are each a product of an arbitrary number of factors, each of which take on the form $\sum_{\vec{p}} g(\vec{p},i\omega_p)\epsilon_{\vec{p}}^m$, with m equal to zero or one, and with arbitrarily complex frequency dependence of the individual factors.  

We show that just as in the finite U case\cite{Khurana,ZlaticHorvatic}, the optical conductivity is given by the expression
\barray \sigma^{\alpha\beta}(\omega) &=& \frac{2}{i\omega} \sum_{\vec{p},i\omega_p} \G(\vec{p},i\omega_p)v_{\vec{p}}^\alpha v_{\vec{p}}^\beta\times\nn\\
&&[\G(\vec{p},\omega + i\eta+i\omega_p)-\G(\vec{p},i\eta+i\omega_p)],
\label{conductivityintro}
\earray
where $v_{\vec{p}}^\alpha$ is the component of the velocity in the $\alpha$ direction (\disp{current}). We show that this formula can be applied at each order of the $\lambda$ expansion.
 
We show that there is a self consistent mapping between the ECFL theory of the infinite-dimensional \tJ model and the ECFL theory of the infinite-U Anderson impurity model(AIM)\cite{ECFLAM}. This mapping is similar in spirit to the mapping first discussed by Georges and Kotliar for the Hubbard model \cite{GeorgesKotliar}, but is made directly in the infinite $U$ limit here.
 In this mapping, $\GH_{i,i}[\tau_i,\tau_f]$ and $\mu_{i,i}[\tau_i,\tau_f]$ of the \tJ model are mapped to the objects $\GH[\tau_i,\tau_f]$ and $\mu[\tau_i,\tau_f]$ of the Anderson model,
 written with the same symbols, but without the spatial or momentum labels. This mapping is valid under the self-consistency condition
\beq \sum_{\vk} \epsilon_{\vk}\GH(k) = \sum_{\vk}\frac{|V_{\vk}|^2}{i\omega_n - \widetilde{\epsilon_{\vk}}}\GH(i\omega_k), \label{AMmappingintro}\eeq
where $\epsilon_{\vec{k}}$ is the dispersion of the lattice in the \tJ model, and $V_{\vk}$ and $\widetilde{\epsilon_{\vk}}$ are the hybridization and dispersion of the bath respectively in the Anderson impurity model. This self-consistency condition is shown to be equivalent to the standard self-consistency condition from DMFT\cite{GeorgesKotliar,GeorgesKotliaretal}. We also show that the mapping holds to each order in $\lambda$ under the same self-consistency condition. We note that this implies that ECFL computations for the infinite-dimensional \tJ model can be done with a DMFT-like self-consistency loop involving ECFL computations for the AIM. However, since the $\lambda$ expansion provides integral equations which are relatively straightforward to solve numerically, this is not necessary as the \tJ model equations can be solved directly. 
\subsection{Outline of the paper}

The paper is structured as follows. In section \ref{Preliminaries}, some basic facts about lattice sums in the limit of large dimensions and the ECFL equations of motion as well as the $\lambda$ expansion are reviewed. Additionally, the spatial dependence of various standard and ECFL specific objects in the limit of large dimensions is stated. Finally, we introduce a class of local functions denoted class-L functions; these turn out to  play a central role in the ECFL in the limit of large dimensions. In sections \ref{EOMlarged} and \ref{zero source}, Eqs. (\ref{Psiid}) and (\ref{Phiid}) are proven in general and to each order in $\lambda$, and the locality of the Dyson-Mori self energy is shown as a consequence. In section \ref{conductivitylarged}, \disp{conductivityintro} is shown to hold in general and to each order in $\lambda$. In section \ref{integralequations}, the ECFL self-consistent integral equations are derived to $O(\lambda^2)$ in the large-d limit. Finally, in section \ref{Anderson}, the ECFL of the infinite dimensional \tJ model is mapped onto the ECFL of the infinite-U AIM under the self-consistency condition \disp{AMmappingintro}. This is done in general and to each order in $\lambda$. 

\section{Preliminaries\label{Preliminaries}}

\subsection{Spatial dependence of lattice sums in large d dimensions}
We take the hopping to nearest neighbor sites on the d-dimensional hypercube. In this case, it is well known \cite{MetznerVollhardt} that $t_{ij} \to \frac{1}{ \sqrt{2d}} t_0$ with $t_0 $ of $O(1)$. We would like to exploit the smallness of individual  $t_{ij}$'s, these can only contribute (after multiplying with another like object),
 if one of the indices is summed over the d-neighbors  as in the simplest example $\sum_j t^2_{ij} =  t_0^2$. Extending this argument further, for a pair of sites  $(i,m)$ located at a (Manhattan metric) distance $r_{im}$  on the hypercube, suppose there are two objects $W_{i,m}$ and $V_{i,m}$ who both have the dependence on $r_{im}$: $V_{i,m};W_{i,m} \sim O\left(\frac{1}{\left(\sqrt{d}\right)^{ r_{im}}}\right)$. Then it follows that 
 \beq W_{i,\bb{n}}V_{\bb{n},m}  \sim  O\left(\frac{1}{\left(\sqrt{d}\right)^{ r_{im}}}\right). \label{sumorder}\eeq
Here, and in the rest of the paper, bold and repeated indices are summed and/or integrated over. This relation can be understood by first considering the case that the site $\bb{n}$ is on one of the shortest paths between $i$ and $m$. In this case, $r_{i\bb{n}} + r_{\bb{n}m}=r_{im}$ proving the relation. If, $\bb{n}$ is a certain distance $r_o$ off of a shortest path, then $r_{i\bb{n}} + r_{\bb{n}m}=r_{im} + 2 r_o$. This introduces and extra factor of $\frac{1}{d^{r_o}}$ into the lattice sum in \disp{sumorder}.  However, this factor is exactly cancelled by the $d^{r_0}$ choices for the site $\bb{n}$. In this argument, the number of shortest paths between $i$ and $m$ is taken to be $O(1)$.

\subsection{ECFL Equations of Motion and the $\lambda$ expansion}

The ECFL equations of motion for the finite dimensional \tJ model can be found in \refdisp{Monster}. There is some freedom in how these equations are written because one may add terms to them which vanish identically in the exact solution, but play a non-trivial role when implementing approximations (such as the $\lambda$ expansion). We denote the version of these equations with no added terms the minimal theory, and the version containing the added terms the symmetrized theory (since the added terms make the resulting expressions symmetric in a certain sense). In \refdisp{Monster}, the ECFL equations of motion for the symmetrized theory are derived, and the added terms required to go from the minimal theory to the symmetrized theory are singled out. The ECFL equations for the minimal theory, which are the ones used in this paper and in \refdisp{ECFLDMFT}, can therefore be obtained from those in \refdisp{Monster} by dropping these extra terms.

Setting $J\to0$ (as discussed in section \ref{model}), we write the minimal theory ECFL equations of motion in expanded form.
\barray
\GHI[i,m]&=&  ( \chem  - \partial_{\tau_i} - \V_i) \ \delta[i,m] + t[i,m] \ ( 1- \lambda\gamma[i]) +\nn \\
&&  \lambda t[i, \bb{j}] \ \xi^* . \GH[\bb{j}, \bb{n}]. \Lambda_* [ \bb{n}, m; i], \nn \\
\mu[i,m]&=& (1- \lambda\gamma[i]) \delta[i,m] - \lambda t[i, \bb{j}] \ \xi^* . \GH[\bb{j}, \bb{n}]. \U_* [ \bb{n}, m; i], \nn\\
\label{set1}
\earray
where $\V_i\equiv\V_i(\tau_i)$ is the Bosonic Schwinger  source function, and we have used the notation $\delta[i,m]=\delta_{i,m}\delta(\tau_i-\tau_m)$ and $t[i,m]=t_{i,m}\delta(\tau_i-\tau_m)$.
These exact relations give the required objects $\GH$ and $\mu$  in terms of the vertex functions.
Here we also note that the local (in space and time) Green's function $\gamma[i]$, and the vertices $\Lambda[n,m;i]$ and $\U[n,m;i]$, are defined as
\barray
\gamma[i]&=& \mu^{(k)}[\bb{n},i^+] . \GH^{(k)}[i,\bb{n}]; \;\;\;
\Lambda[n,m;i]= - \frac{\delta}{\delta \V_i} \GHI[n,m];  \nn\\
&& \U[n,m;i]=  \frac{\delta}{\delta \V_i} \mu[n,m], 
 \label{set2}
\earray
 where we have used the notation $M^{(k)}_{\sigma_1,\sigma_2}= \sigma_1 \sigma_2 M_{\bar{\sigma}_2,\bar{\sigma}_1}$ to denote the time reversed  matrix $M^{(k)}$ of an arbitrary matrix $M$. These exact relations give the vertex functions    in terms of the objects $\GH$ and $\mu$.
The vertices defined above ($\Lambda$ and $\U$) have four spin indices, those of the object being differentiated and those of the source. For example, $\U^{\sigma_1\sigma_2}_{\sigma_a\sigma_b}[n,m;i]=  \frac{\delta}{\delta \V^{\sigma_a\sigma_b}_i} \mu_{\sigma_1\sigma_2}[n,m]$. In \disp{set1}, $\xi_{\sigma_a\sigma_b}=\sigma_a\sigma_b$, and the $^*$ indicates that these spin indices should also be carried over (after being flipped) to the bottom indices of the vertex, which is also marked with a $_*$. The top indices of the vertex are given by the usual matrix multiplication. An illustrative example is useful here:  $\left(\xi^* . \GH[j, \bb{n}]. \U_* [ \bb{n}, m; i]\right)_{\sigma_1\sigma_2}= \sigma_1\sigma_{\bb{a}}  \ \GH_{\sigma_{\bb{a}},\sigma_{\bb{b}}}[j, \bb{n}]\frac{\delta}{\delta \V^{\bar{\sigma}_1\bar{\sigma}_{\bb{a}}}_i }\mu_{\sigma_{\bb{b}},\sigma_2}[\bb{n},m] $. Finally, in order to ensure that the shift identities (\refdisp{Monster}) are satisfied, the substitution $t_{ij} \to t_{ij} + \frac{u_0}{2}\delta_{ij}$ is made, where $u_0$ is the second chemical potential. For the sake of clarity, this substitution will be ignored in the proofs given below, although they are easily generalized to account for it. This generalization is discussed at the end of section \ref{EOMlarged}.

The $\lambda$ expansion is obtained by expanding \disp{set1} and \disp{set2} iteratively in the continuity parameter $\lambda$. The $\lambda=0$ limit of these equations is the free Fermi gas. Therefore, a direct expansion in $\lambda$ will lead to a series in $\lambda$ in which each term is made up of the hopping $t_{ij}$ and the free Fermi gas Green's function $\GH_0[i,f]$. As is the case in the Feynman series, this can be reorganized into a skeleton expansion in which only the skeleton graphs are kept and $\GH_0[i,f] \to \GH[i,f]$. However, one can also obtain the skeleton expansion directly by expanding \disp{set1} and \disp{set2} in $\lambda$, but treating $\GH[i,f]$ as a zeroth order (i.e. unexpanded) object in the expansion. This expansion is carried out to second order for the finite-dimensional case in \refdisp{Monster}. In doing this expansion, one must evaluate the functional derivative $\frac{\delta \GH}{\delta \V}$. This is done with the help of the following useful formula which stems from the product rule for functional derivatives.
\beq  \frac{\delta \GH[i,m]}{\delta \V_r} = \GH[i,\bb{x}].\Lambda[\bb{x},\bb{y},r].\GH[\bb{y},m]. \label{Derivative}\eeq
This is an exact formula and will be used extensively in the arguments given below. Within the $\lambda$ expansion, the LHS is evaluated to a certain order in $\lambda$ by taking the vertex $\Lambda$ on the RHS to be of that order in $\lambda$.

\subsection {Leading order spatial dependence of various objects\label{leadspat}}
All objects may be expanded in the inverse square root of the number of spatial dimensions d. The lowest order term in the physical Green's function $\G[i,f]$ must be at least $O\left(\frac{1}{\left(\sqrt{d}\right)^{r_{if}}}\right)$. This must be so because it takes at least $r_{if}$ hops to get from the site $i$ to the site $f$. Any terms that contribute to $\G[i,f]$ at higher order than $O\left(\frac{1}{\left(\sqrt{d}\right)^{r_{if}}}\right)$ are neglected in the large d limit. In a similar vain, the lowest order term in $\GH[i,f]$, $\GHI[i,f]$, $\mu[i,f]$, $\Lambda[i,f;r]$, and $\U[i,f;r]$ must be at least $O\left(\frac{1}{\left(\sqrt{d}\right)^{r_{if}}}\right)$. Furthermore, using the real space version of \disp{Ggmu} and \disp{sumorder}, we see that any terms of higher order than this in $\GH[i,f]$ and $\mu[i,f]$ will result in a higher order term in $\G[i,f]$ and may therefore be neglected as well. Finally, using matrix inversion in the space-time indices, we see that higher order terms may also be dropped from $\GHI[i,f]$ as these will lead to higher order terms in $\GH[i,f]$, and using \disp{set1}, higher order terms may be dropped from $\Lambda[i,f;r]$, and $\U[i,f;r]$ as these will lead to higher order terms in $\GHI[i,f]$ and $\mu[i,f]$ respectively. In summary, in all objects:  $\G[i,f]$, $\GH[i,f]$, $\GHI[i,f]$, $\mu[i,f]$, $\Lambda[i,f;r]$, and $\U[i,f;r]$, terms of higher order than $O\left(\frac{1}{\left(\sqrt{d}\right)^{r_{if}}}\right)$ may be neglected in the large d limit. 

We also note that the correlation function $\Pi_{\alpha\beta}[i,f]$ appearing in \disp{Pi} must be at least $O\left(\frac{1}{d^{r_{if}}}\right)$. This is due to the fact that unlike the creation and destruction operators which appear in the Green's function, the current operators appearing in this correlation function conserve particle number. Hence, one must hop from site $i$ to site $f$ and back, which takes $2 \times r_{if}$ hops. Any terms that contribute to $\Pi_{\alpha\beta}[i,f]$ at higher order than $O\left(\frac{1}{d^{r_{if}}}\right)$ are neglected in the large d limit.

\subsection{Class L functions\label{classL}}

For the  arguments given below, we need to define a class of {\em localized functions},  denoted as class L functions. A class L function $L_i$ has three properties. 
\begin{itemize}
\item (a) $L_i \sim O\left(\frac{1}{d^0}\right)$.
\item  (b) $L_i$ is a function of only one site $i$, and an arbitrary number of time variables. Upon turning off the sources, it becomes translationally invariant, but an arbitrary function of frequencies.
\item  (c) The $\V$ source derivative of $L_i$ is also localized:
\beq
 \frac{\delta}{\delta \V_i} L_j = \delta_{ij} L'_i,
\eeq
with $L'_i$ again a Class-L function. 
\end{itemize}
Our proofs deal with functions that turn out to be of this class. Iterating property (c), the following equation must hold for any positive integer $s$.
\beq \frac{\delta}{\delta\V_{r_1}}\ldots\frac{\delta}{\delta\V_{r_s}}L_i = \delta_{ir_1}\ldots\delta_{ir_s}\frac{\delta}{\delta\V_i(\tau_{r_1})}\ldots\frac{\delta}{\delta\V_i(\tau_{r_s})}L_i.  \nn \\
\label{classLloc}\eeq  
In the presence of the current source $\kappa$ ( \disp{currentsource}), class L functions acquire one additional property (d): Consider a typical contribution to $\Pi_{\alpha\beta}[i,f]$ ( \disp{PiG}) denoted by $O_{if}$
\beq O_{if} = W_{f,\bb{x}} \  \frac{\delta}{\delta \kappa_i^\alpha} \left(L_{\bb{x}}\right) \  V_{\bb{x},f}, \eeq
where the functions  $V_{\bb{x},f}, W_{f,\bb{x}}\sim O\left(\frac{1}{\left(\sqrt{d}\right)^{ r_{\bb{x}f}}}\right)$. Then, neglecting terms of higher order than $O\left(\frac{1}{d^{r_{if}}}\right)$ in $O_{if}$, $\sum_{i-f}O_{if}\to0$ as $\A \to 0$.   Again iterating property (c) and using property (d), the following must hold for any nonnegative integer $s$:
\beq \!\!\!\! \sum_{i-f}\left(W_{f,\bb{x}} \ \frac{\delta}{\delta \kappa_i^\alpha}\frac{\delta}{\delta\V_{\bb{x}}(\tau_{r_1})}\ldots\frac{\delta}{\delta\V_{\bb{x}}(\tau_{r_s})}\left(L_{\bb{x}}\right) \  V_{\bb{x},f} \right)_{ \A \to 0} = 0. \nn \\
\!\! \label{derkapzero}\eeq

\section{Limit of Large dimensionality through the ECFL  equations of motion}

\subsection{Simplification of the ECFL self energies \label{EOMlarged}}
We use notation in which we indicate spatial dependence by subscripts, so that $\GH[i,j]\to \GH_{i,j}[\tau_i,\tau_j]$, 
and recall that $t[i,j]= t_{i,j} \ \delta(\tau_i-\tau_j)$, $\delta[i,j]=\delta_{i,j} \ \delta(\tau_i-\tau_j)$, and $\delta[\tau_i,\tau_j]= \delta(\tau_i-\tau_j)$ etc.
After some inspection of \disp{set1} and \disp{set2} in the limit of high dimension, we make an Ansatz - to be proven below - namely
\barray
\GHI[i,m] &=&  ( \chem  - \partial_{\tau_i} - \V_i) \ \delta[i,m] + \ t[i,m] \ ( 1- \lambda\gamma[i]) \nn\\
&& - \lambda \ \delta_{i,m} \ \chi_i[\tau_i,\tau_m] + \lambda \  t_{i,m} \ \Psi_i[\tau_i,\tau_m],  \nn \\
\mu[i,m] &=&  \delta[i,m] (1- \lambda\gamma[i]) +  \lambda \ \delta_{i,m} \ \Psi_i[\tau_i,\tau_m], \label{local-ansatzginmu}
\earray
where $ \Psi_i[\tau_i,\tau_m] $, $ \chi_i[\tau_i,\tau_m] $, and $\gamma[i]$ are class L functions. We will prove \disp{local-ansatzginmu} by assuming that it is true, and then showing that this assumption is consistent with the equations of motion (Eqs. (\ref{set1}) and (\ref{set2})). This argument will consist of a loop which begins with \disp{local-ansatzginmu}. Then, substituting this equation into \disp{set2}, we will derive a certain form for $\Lambda$, $\U$, and $\gamma$. Finally, substituting these objects into \disp{set1}, and using simplifications which occur in the large d limit, we will complete the loop and arrive back at \disp{local-ansatzginmu}. 

Substituting \disp{local-ansatzginmu} into \disp{set2}, we find that the vertices and $\gamma[i]$ have the following form.
\barray
&&\Lambda[n,m;i] = \delta_{i,n} \delta_{i,m} \ A_i[\tau_n,\tau_m;\tau_i] + \delta_{i,n} t_{n,m} \ B_i[\tau_n,\tau_m;\tau_i], \nn \\
&&\U[n,m;i] = - \delta_{i,n} \delta_{i,m} \ B_i[\tau_n,\tau_m;\tau_i], \nn\\
&&\gamma[i] = \left(1- \lambda\gamma^{(k)}[i]\right)\GH^{(k)}[i,i] +  \lambda \ \Psi_i^{(k)}[\tau_{\bb{j}},\tau_i]\GH_{ii}^{(k)}[\tau_i,\tau_{\bb{j}}], \nn\\
\label{local-ansatz}
\earray
where we defined two new functions:
\barray
A_i[\tau_n,\tau_m;\tau_i]&=&  \delta[\tau_i,\tau_n] \delta[\tau_i,\tau_m] \ \iden +  \lambda \ \frac{\delta}{\delta \V_i} \chi_i[\tau_n,\tau_m], \nn \\
B_i[\tau_n,\tau_m;\tau_i]&=& \lambda \ \delta[\tau_n,\tau_m] \frac{\delta}{\delta \V_i} \gamma_i[\tau_n]  -  \lambda \ \frac{\delta}{\delta \V_i} \Psi_i[\tau_n,\tau_m]. \nn\\
\label{eq-A}
\earray
Here $A_i$ and $B_i$ are class L functions since they inherit this property from $\Psi_i$, $\chi_i$, and $\gamma[i]$ by functional differentiation. Substituting \disp{local-ansatz} into \disp{set1} and comparing with \disp{local-ansatzginmu}, 
\barray
\chi_i[\tau_i,\tau_m]&=&   - \  t_{i,\bb{j}} \ \xi^*. \ \GH_{\bb{j}, i} [\tau_i, \tau_{\bb{n}}] . \ A_{i, *}[ \tau_{\bb{n}},\tau_m; \tau_i], \nn\\
 \Psi_i[\tau_i,\tau_m]&=& \  t_{i,\bb{j}} \ \xi^*. \ \GH_{\bb{j}, i} [\tau_i, \tau_{\bb{n}}] . \ B_{i, *}[ \tau_{\bb{n}},\tau_m; \tau_i]. \label{eq-F}
\earray
 If we can now show that $\chi_i$, $\Psi_i$, and $\gamma[i]$ as defined in \disp{local-ansatz} and \disp{eq-F} are Class L functions, we will have justified our Ansatz and therefore we will have proven all of the above equations. To do this, we must show that $\GH_{ii}[\tau_i,\tau_m]$ and  $t_{i,\bb{j}}  \ \GH_{\bb{j}, i} [\tau_i, \tau_m]$ are Class L functions. Taking their functional derivatives we obtain:
\barray 
\frac{\delta}{\delta \V_r} t_{i,\bb{j}}  \ \GH_{\bb{j}, i} [\tau_i, \tau_m] = t_{i,\bb{j}}\GH_{\bb{j}, r}[\tau_i, \tau_{\bb{k}}] \ A_r[ \tau_{\bb{k}},\tau_{\bb{l}}; \tau_r]  \GH_{r, i}[\tau_{\bb{l}}, \tau_m] \nn\\
 + t_{i,\bb{j}}\GH_{\bb{j}, r}[\tau_i, \tau_{\bb{k}}] \ B_r[ \tau_{\bb{k}},\tau_{\bb{l}}; \tau_r]t_{r,\bb{l}} \GH_{\bb{l}, i}[\tau_{\bb{l}}, \tau_m], \nn \\
\label{inductstep} 
\earray
and 
\barray
\frac{\delta}{\delta \V_r}  \GH_{i, i} [\tau_i, \tau_m] &=&\GH_{i, r}[\tau_i, \tau_{\bb{k}}] \ A_r[ \tau_{\bb{k}},\tau_{\bb{l}}; \tau_r]  \GH_{r, i}[\tau_{\bb{l}}, \tau_m]\nn\\
&& +\GH_{i, r}[\tau_i, \tau_{\bb{k}}] \ B_r[ \tau_{\bb{k}},\tau_{\bb{l}}; \tau_r]t_{r,\bb{l}} \GH_{\bb{l}, i}[\tau_{\bb{l}}, \tau_m]. \nn\\
\label{inductstep2} 
\earray
Using \disp{sumorder}, the terms on the RHS of \disp{inductstep} and \disp{inductstep2} survive the large d limit if and only if $r=i$. Moreover, upon making the substitution $r \to i$, we see that the RHS is made up of the same objects that appear on the LHS of the equations (as well as the class L functions $A$ and $B$). Therefore, this argument can be iterated to any number of derivatives acting on $t_{i,\bb{j}}  \ \GH_{\bb{j}, i} [\tau_i, \tau_m]$ or  $\GH_{i, i} [\tau_i, \tau_m]$ (as required by \disp{classLloc}), which are therefore class L functions. Thus, we have shown the self-consistency of our ansatz \disp{local-ansatzginmu}.

The above results hold for any value of $\lambda$, since the proof  was done with $\lambda$ present in all of the equations. In the bare expansion, this would imply that they also hold to each order in $\lambda$. However, this line of reasoning is not as straightforward in the skeleton expansion because each order in the skeleton expansion contains contributions from all orders in the bare expansion. Nonetheless, the above results do hold to each order in $\lambda$ in the skeleton expansion. In proving this, we shall shed more light on the nature of the objects $\Psi_i$, $\chi_i$, $\gamma[i]$, $A_i$, and $B_i$. In particular, we will show that they satisfy a certain explicit form stated below in \disp{objects}. We will do this using an inductive argument, in which we will assume that they have this form through a certain order in $\lambda$, and then substituting this form into the equations of motion, will show that it must hold for the next order. 

 { We  now use the symbol} $R_i$ {\bf as a proxy} for either of the two functions $\GH_{i,i}[\tau_n,\tau_m]$ or $t_{i,\bb{j}}\GH_{\bb{j},i}[\tau_n,\tau_m]$ where the time indices are arbitrary. {\bf Inductive hypothesis}: Through $n^{th}$ order in $\lambda$, \disp{local-ansatzginmu} and \disp{local-ansatz} hold. Through $n-1^{st}$ order in $\lambda$, the objects $\Psi_i$, $\chi_i$, and $\gamma[i]$, and through $n^{th}$ order, the objects $A_i$ and $B_i$, (all denoted below by the generic object $L_i$) can be written as the following product (multiplied by some delta functions in time variables): 
\beq \left(L_i\right)^{(n)} = \lambda^n (R_i)^m \label{objects}, \eeq
where $m$ is arbitrary. We first examine the base case of zeroth order. In this case,
\barray
A^{(0)}_i[\tau_n,\tau_m;\tau_i] =  \delta[\tau_i,\tau_n] \delta[\tau_i,\tau_m] ; \;\;\;
B^{(0)}_i[\tau_n,\tau_m;\tau_i] =0. \nn\\
\label{basecase}
\earray
Clearly the hypothesis is satisfied. Now, we prove the inductive step. Explicitly displaying the order  in $\lambda$ of all objects, the equations for $\chi$, $\Psi$, and $\gamma$ (Eqs. (\ref{eq-F}) and (\ref{local-ansatz})) become
\barray
\chi^{(n)}_i[\tau_i,\tau_m]&=& - \ t_{i,\bb{j}} \ \xi^*. \ \GH_{\bb{j}, i} [\tau_i, \tau_{\bb{n}}] . \ A^{(n)}_{i, *}[ \tau_{\bb{n}},\tau_m; \tau_i], \nn\\
\Psi^{(n)}_i[\tau_i,\tau_m]&=& \ t_{i,\bb{j}} \ \xi^*. \ \GH_{\bb{j}, i} [\tau_i, \tau_{\bb{n}}] . \ B^{(n)}_{i, *}[ \tau_{\bb{n}},\tau_m; \tau_i], \nn\\
\gamma^{(n)}[i] &=& - \lambda \ \gamma^{(k)(n-1)}[i]\GH^{(k)}[i,i] \nn\\
&& +  \lambda  \ \Psi_i^{(k)(n-1)}[\tau_{\bb{j}},\tau_i]\GH_{ii}^{(k)}[\tau_i,\tau_{\bb{j}}].\nn\\
\label{eq-Clambda}
\earray
By the inductive hypothesis, $\chi^{(n)}_i$, $\Psi^{(n)}_i$, and $\gamma^{(n)}[i]$ have the required form. The equations for $A$ and $B$ (\disp{eq-A}) become 
\barray
A^{(n+1)}_i[\tau_n,\tau_m;\tau_i] &=&   \lambda \ \left(\sum_{r\leq n}\frac{\delta}{\delta \V_i} \chi^{(r)}_i[\tau_n,\tau_m]\right)^{(n)}, \nn\\
B^{(n+1)}_i[\tau_n,\tau_m;\tau_i] &=& \lambda \ \delta[\tau_n,\tau_m] \left(\sum_{r\leq n}\frac{\delta}{\delta \V_i} \gamma^{(r)}_i[\tau_n]\right)^{(n)}  \nn\\
&&-  \lambda \ \left(\sum_{r\leq n}\frac{\delta}{\delta \V_i} \Psi^{(r)}_i[\tau_n,\tau_m]\right)^{(n)}.\nn\\
\label{eq-Alam} 
\earray 
To see that $A^{(n+1)}$ and $B^{(n+1)}$ have the required form we note that for all $l \leq n$, 
\barray	
&&\left(\frac{\delta}{\delta \V_r} t_{i,\bb{j}}  \ \GH_{\bb{j}, i} [\tau_i, \tau_m]\right)^{(l)} = \nn\\
&& t_{i,\bb{j}}\GH_{\bb{j}, r}[\tau_i, \tau_{\bb{k}}] \ A^{(l)}_r[ \tau_{\bb{k}},\tau_{\bb{l}}; \tau_r]  \GH_{r, i}[\tau_{\bb{l}}, \tau_m] \nn\\
&&+ t_{i,\bb{j}}\GH_{\bb{j}, r}[\tau_i, \tau_{\bb{k}}] \ B^{(l)}_r[ \tau_{\bb{k}},\tau_{\bb{l}}; \tau_r]t_{r,\bb{l}} \GH_{\bb{l}, i}[\tau_{\bb{l}}, \tau_m], \nn\\
\label{inductsteplam} 
\earray	
and
\barray
\left(\frac{\delta}{\delta \V_r}  \GH_{i, i} [\tau_i, \tau_m]\right)^{(l)}  =\GH_{i, r}[\tau_i, \tau_{\bb{k}}] \ A^{(l)} _r[ \tau_{\bb{k}},\tau_{\bb{l}}; \tau_r]  \GH_{r, i}[\tau_{\bb{l}}, \tau_m]\nn\\
+\GH_{i, r}[\tau_i, \tau_{\bb{k}}] \ B^{(l)} _r[ \tau_{\bb{k}},\tau_{\bb{l}}; \tau_r]t_{r,\bb{l}} \GH_{\bb{l}, i}[\tau_{\bb{l}}, \tau_m].\nn\\
 \label{inductstep2lam} 
\earray
In the limit of large dimensions, $r\to i$. We can therefore (using the inductive hypothesis) write the RHS of \disp{inductsteplam} and \disp{inductstep2lam} as $\lambda^l(R_i)^m$. Applying \disp{objects} (which has been shown to hold for $\chi^{(n)}_i$, $\Psi^{(n)}_i$, and $\gamma^{(n)}[i]$) to \disp{eq-Alam}, we may write
\barray
A^{(n+1)}_i &=& \sum_{r=0}^{n}\lambda^{r+1}\left(\frac{\delta}{\delta \V_i}(R_i)^m\right)^{(n-r)}, \nn\\
B^{(n+1)}_i &=&  \sum_{r=0}^{n}\lambda^{r+1}\left(\frac{\delta}{\delta \V_i}(R_i)^m\right)^{(n-r)}.\nn\\
\label{ABnpone}   
\earray
\disp{ABnpone}, in conjunction with \disp{inductsteplam} and \disp{inductstep2lam}, shows that $A^{(n+1)}_i $ and $B^{(n+1)}_i $ have the required form. This completes the proof.

{ Since $t_{i,j}$ is independent of the source, the substitution $t_{i,j} \to t_{i,j} + \frac{u_0}{2}\delta_{i,j}$ can be made directly into all of the above equations. The only problem that could potentially arise involves Eqs. (\ref{inductstep}) and (\ref{inductstep2}),  where the large d simplifications are actually  used. However, one can check that  this substitution  does not affect the simplifications. Therefore, this substitution merely adds the term $\lambda\frac{u_0}{2}\delta_{i,m} \Psi_i[\tau_i,\tau_m] - \lambda\frac{u_0}{2}\delta[i,m]\gamma[i]$ to $\GHI[i,m]$, and everywhere replaces the local function $t_{i,\bb{j}}\GH_{\bb{j},i}[\tau_n,\tau_m]$ with the local function  $t_{i,\bb{j}}\GH_{\bb{j},i}[\tau_n,\tau_m] + \frac{u_0}{2}\GH_{i,i}[\tau_n,\tau_m] $. This can be seen explicitly in the $O(\lambda^2)$ equations in section \ref{integralequations}, and does not change the general structure of the solution.}

\subsection{The zero source limit\label{zero source}}
Setting the sources to zero, the system becomes translationally invariant so that all objects can be written in momentum space. Additionally, $\gamma[i] \to \frac{n}{2}$. Then, the above results can be summed up in the following formulae (in which we set $\lambda=1$):
\barray
\GHI(k) &=& i\omega_k + \chem - \varepsilon_k(1-\frac{n}{2})-\chi(i\omega_k) - \varepsilon_k \Psi(i\omega_k), \nn\\
\mu(k) &=& 1-\frac{n}{2} +\Psi(i\omega_k),
\label{momindself}
\earray
where $\Psi(i\omega_k)$ and $\chi(i\omega_k)$ are the two momentum independent self-energies of the ECFL in infinite dimensions. In terms of these self-energies, the physical Green's function is written as
\beq
\G(k) = \frac{ 1-\frac{n}{2} +\Psi(i\omega_k)}{i\omega_k + \chem - \varepsilon_k(1-\frac{n}{2})-\chi(i\omega_k) - \varepsilon_k \Psi(i\omega_k)}.
\eeq
Comparing with the standard form of the Green's function in terms of the Dyson-Mori self energy
\beq \G(k) = \frac{1-\frac{n}{2}}{i\omega_k + \chem-\epsilon_k(1-\frac{n}{2})-\Sigma_{DM}(k)}, \eeq
we see the momentum independence of the  ECFL self energies
 $\Sigma_{DM}(k) =\Sigma_{DM}(i\omega_k)$, and 
\beq \!\!\!
\Sigma_{DM}(i\omega_k)= \frac{(i\omega_k + \mbox{\boldmath{$\mu$}})\Psi(i\omega_k)+(1-\frac{n}{2})\chi(i\omega_k)}{1-\frac{n}{2}+\Psi(i\omega_k)}. \nn \\ \label{selfrelatemin} \! \eeq

\subsection{Conductivity in the limit of large dimensions\label{conductivitylarged}}
It is well known that for the finite-U Hubbard model in the limit of large dimensions, for zero wave vector, vertex corrections can be neglected in the current current correlation function \cite{Khurana,GeorgesKotliaretal}. This simple observation allows one to express the optical conductivity in terms of the single particle Green's function as in \disp{sigmaGG}. We show that this is also the case for the infinite dimensional \tJ model. Moreover, a question of practical importance for the purpose of calculating the optical conductivity within the framework of ECFL, is whether or not \disp{sigmaGG} can be applied at each order in the $\lambda$ expansion (as is done in \refdisp{ECFLDMFT}). We show that it can be applied and is the correct procedure. First, we define the relevant objects. 

The Schr\"{o}dinger picture current operator for site j in the direction $\alpha$ is defined as:
\barray 
J_j^{\alpha} &=& i \sum_{k\sigma}v^\alpha_{k,j}X_k^{\sigma 0}X_j^{0\sigma} ; \;\;\; 
v^\alpha_{k,j} = t_{k,j}(\vec{R}_k - \vec{R}_j)_{\alpha},\nn\\
\label{current} 
\earray
so that $v$ is a velocity.
Using the notation ${J}^\alpha[i]=J_i^\alpha(\tau_i) $; $\widetilde{J}^\alpha[i] = J^\alpha[i] - \langle J^\alpha[i]\rangle$, we define the correlation function $\Pi_{\alpha\beta}[i,f]$ and its Fourier transform as 
\barray 
&&\Pi_{\alpha\beta}[i,f] = \langle T_\tau\widetilde{J}^\alpha[i] \widetilde{J}^\beta[f]\rangle ; \nn\\
&&\Pi_{\alpha\beta}(\vec{q},i\Omega_n)  = \int_0^\beta d(\tau_i-\tau_f) \  e^{i\Omega_n(\tau_i-\tau_f) } \nn\\
&& \:\;\;\;\;\;\;\;\;\;\;\;\;\;\;\;\;\;\;\;\;\;\;\;\;\; \times \sum_{i-f} e^{-i\vec{q}\cdot (\vec{R}_i-\vec{R}_f)} \Pi_{\alpha\beta}[i,f].\nn\\
\label{Pi} 
\earray
The optical conductivity can be given in terms of this object as
\beq \sigma^{\alpha\beta}(\omega) = \frac{1}{i\omega - \eta}\left[ \Pi_{\alpha\beta}(\vec{0},\omega+i\eta)-\Pi_{\alpha\beta}(\vec{0},i\eta)\right], \label{conductivity}\eeq
where $\eta=0^+$.
We would like to express the object $\Pi_{\alpha\beta}[i,f]$ as a functional derivative of the Green's function. To this end, we add a source which couples to the current operator
\beq \A \to \A + \sum_{j\alpha}\int_0^\beta d\tau \kappa_j^\alpha(\tau)J_j^\alpha(\tau). \label{currentsource}\eeq
In terms of the $\kappa$ source derivative of the Green's function, and using the definitions $v^\alpha[i,j] = v^\alpha_{i,j}\delta(\tau_i-\tau_j)$;$\kappa^\alpha_i=\kappa^\alpha_i(\tau_i)$, $\Pi_{\alpha\beta}[i,f]$ is given as
\beq  \Pi_{\alpha\beta}[i,f] = -i  \ Tr\left(\frac{\delta}{\delta\kappa^\alpha_i}\G[f,\bb{j}] \ v^\beta[\bb{j},f^+]\right)_{ \A \to 0}, \eeq
where the trace is over the spin degrees of freedom only. We expand the RHS of this equation using \disp{Derivative} (which holds equally well for the $\kappa$ source derivative), finally obtaining an expression for $\Pi_{\alpha\beta}[i,f]$ in terms of the $\kappa$ source derivatives of $\GHI$ and $\mu$. 
\barray
 &&\Pi_{\alpha\beta}[i,f] = \nn\\
 &&i \ Tr\left(\GH[f,\bb{x}] \ \frac{\delta}{\delta\kappa^\alpha_i}\GHI[\bb{x},\bb{y}] \ \GH[\bb{y},\bb{k}]\mu[\bb{k},\bb{j}] \ v^\beta[\bb{j},f^+]\right)_{ \A \to 0} \nn\\
 &&  -i  \ Tr\left(\GH[f,\bb{k}]\frac{\delta}{\delta\kappa^\alpha_i}\mu[\bb{k},\bb{j}] \ v^\beta[\bb{j},f^+]\right)_{ \A \to 0}. 
 \label{PiG}
 \earray

We now consider how the additional source \disp{currentsource} affects the ECFL equations of motion (\disp{set1} and \disp{set2}). The source enters into the equations of motion in the same way as the Hamiltonian does, via its commutator with the destruction operator, $X_i^{0\sigma}$. Moreover, the source has the same form as the Hamiltonian, with the hopping in the kinetic energy replaced by the velocity in the current operator. Therefore, the additional source affects the equations of motion only through the substitution 
\beq t[i,f] \to t[i,f] - i \sum_\alpha \kappa_f^\alpha \  v^\alpha[i,f]. \label{subt}\eeq
Thus, the new equations of motion can be read off from \disp{set1} as
\barray
\GHI[i,m]&=&  ( \chem  - \partial_{\tau_i} - \V_i) \ \delta[i,m] + \nn\\
&&(t[i,m]- i \sum_\alpha \kappa_m^\alpha \  v^\alpha[i,m]) \ ( 1- \lambda\gamma[i]) +\nn \\
&&  \lambda (t[i, \bb{j}] - i \sum_\alpha \kappa_{\bb{j}}^\alpha \  v^\alpha[i,\bb{j}]) \ \xi^* . \GH[\bb{j}, \bb{n}]. \Lambda_* [ \bb{n}, m; i], \nn \\
\mu[i,m]&=& (1- \lambda\gamma[i]) \delta[i,m] -\nn\\
&& \lambda (t[i, \bb{j}]- i \sum_\alpha \kappa_{\bb{j}}^\alpha \  v^\alpha[i,\bb{j}])\ \xi^* . \GH[\bb{j}, \bb{n}]. \U_* [ \bb{n}, m; i]. \nn\\
\label{set1conduct}
\earray
Since there is no source derivative with respect to $\kappa$ in the equations of motion and $v^\alpha[i,f]$ is of the same order in $\frac{1}{\sqrt{d}}$ as $t[i,f]$, all of the results derived in section \ref{EOMlarged} continue to hold after making the substitution in \disp{subt}. In particular, we showed that $\GHI[i,m] $ and $\mu[i,m]$ have the following form (\disp{local-ansatzginmu}).
\barray
\GHI[i,m] &=&  ( \chem  - \partial_{\tau_i} - \V_i) \ \delta[i,m] -\lambda \ \delta_{i,m} \ \chi_i[\tau_i,\tau_m] + \nn\\
&&( t[i,m] - i \sum_\alpha \kappa_m^\alpha \ v^\alpha[i,m] ) \ ( 1- \lambda\gamma[i])  \nn\\
 &&+ \lambda \  (t_{i,m} - i \sum_\alpha \kappa_m^\alpha \  v^\alpha_{i,m} ) \ \Psi_i[\tau_i,\tau_m],  \nn \\
\mu[i,m] &=&  \delta[i,m] (1- \lambda\gamma[i]) +  \lambda \ \delta_{i,m} \ \Psi_i[\tau_i,\tau_m],
\label{local-ansatzginmukappa}
\earray
where $\chi_i$, $\Psi_i$, and $\gamma[i]$ have properties (a)-(c) of class L functions (sec.\ref{classL}), and are defined by Eqs. (\ref{local-ansatzginmu}) through (\ref{eq-F}). We shall now further assume that they also satisfy property (d) (\disp{derkapzero}) and show that this assumption is consistent with their definitions. This, in turn, will allow us to demonstrate the validity of \disp{sigmaGG}.

Our task is then to show that $\chi_i$, $\Psi_i$, and $\gamma[i]$, as defined in the last line of \disp{local-ansatz} and \disp{eq-F}, satisfy \disp{derkapzero}. By \disp{eq-A}, $A_i$ and $B_i$ satisfy \disp{derkapzero} since they inherit this property from $\chi_i$, $\Psi_i$, and $\gamma[i]$. It remains to show that $\GH_{x,x}[\tau_n,\tau_m]$ and  $(t_{x,\bb{j}} -i\sum_\alpha\kappa^\alpha_{\bb{j}}(\tau_n)v^\alpha_{x,\bb{j}})   \ \GH_{\bb{j}, x} [\tau_n, \tau_m]$ (the time indices are arbitrary) satisfy this equation.  

Defining the notation $w_{i,f}(\tau_i) \equiv  t_{i,f} - i \sum_\alpha \kappa_f^\alpha(\tau_i) \  v^\alpha_{i,f}$, and using (the $\kappa$ source derivative version of) \disp{Derivative} as well as \disp{local-ansatzginmukappa}, we find that
\begin{widetext}
\barray
&&\!\!\!\!\! \left(\frac{\delta}{\delta\kappa_i^\alpha} w_{x,\bb{j}}(\tau_n) \ \GH_{\bb{j}, x} [\tau_n, \tau_m]\right)_{\A\to0}= -i\delta[\tau_i,\tau_n]v^\alpha_{x,i} \ \GH_{i,x} [\tau_i, \tau_m]+ \ it_{x,\bb{j}}\GH_{\bb{j},\bb{a}}[\tau_n,\tau_{\bb{a}}](1-\lambda\gamma[\bb{a}]\delta[\tau_{\bb{a}},\tau_i] +\lambda\Psi_{\bb{a}}[\tau_{\bb{a}},\tau_i])v^\alpha_{\bb{a},i}g_{i,x}[\tau_i,\tau_m] \nn\\
&& +\lambda \ t_{x,\bb{j}}\GH_{\bb{j},\bb{a}}[\tau_n,\tau_{\bb{a}}]\frac{\delta}{\delta\kappa_i^\alpha}\left(\gamma[\bb{a}]\delta[\tau_{\bb{a}},\tau_{\bb{b}}]-\Psi_{\bb{a}}[\tau_{\bb{a}},\tau_{\bb{b}}]\right)t_{\bb{a},\bb{b}}g_{\bb{b},x}[\tau_{\bb{b}},\tau_m] + \lambda \ t_{x,\bb{j}}\GH_{\bb{j},\bb{a}}[\tau_n,\tau_{\bb{a}}]\frac{\delta}{\delta\kappa_i^\alpha}\left(\chi_{\bb{a}}[\tau_{\bb{a}},\tau_{\bb{b}}]\right)g_{\bb{a},x}[\tau_{\bb{b}},\tau_m],
\label{inductstepkappa}
\earray
\end{widetext}
where the RHS is also evaluated in the $\A \to 0$ limit. We now substitute  this into \disp{derkapzero} (with $s=0$). The last two terms must vanish by assumption (where $\bb{a}$ has taken the place of $\bb{x}$). The first term contains two paths from $i$ to $f$, both via $\bb{x}$. Hence, this term must vanish in the large d limit unless $\bb{x}=i$ or $\bb{x}=f$. The former also vanishes since $v^\alpha_{i,i}=0$ while the latter must vanish due to the sum over $i-f$ and the odd parity of $v^\alpha_{i,f}$. The same reasoning applies to the second term except that in this term the $\bb{x}=i$ case vanishes by the odd parity of $v^\alpha_{i,f}$. Hence, we have shown that $(t_{x,\bb{j}} -i\sum_\alpha\kappa^\alpha_{\bb{j}}(\tau_n)v^\alpha_{x,\bb{j}})   \ \GH_{\bb{j}, x} [\tau_n, \tau_m]$ satisfies \disp{derkapzero} with $s=0$. A completely analogous argument shows that this is also the case for $\GH_{x,x}[\tau_n,\tau_m]$. Using \disp{inductstep} and \disp{inductstep2} (in particular the fact that the RHS is made up of the same objects as the LHS), the above argument can be used to show that the result holds for any value of s. Thus, we have demonstrated the self-consistency of our ansatz (\disp{derkapzero}).

Substituting \disp{local-ansatzginmukappa} into \disp{PiG}, and using \disp{derkapzero}, we find that
\beq  \sum_{i-f}\Pi_{\alpha\beta}[i,f] =   \sum_{i-f}Tr\left(\G[f,\bb{k}]v^\alpha[\bb{k},i]\G[i,\bb{j}]  \ v^\beta[\bb{j},f^+]\right)_{ \A \to 0}.\nn\\
 \label {PiGwGw}\eeq
Substituting this equation into \disp{conductivity}, the optical conductivity may be expressed as
\barray \sigma^{\alpha\beta}(\omega) &=& \frac{2}{i\omega} \sum_{\vec{p},i\omega_p} \G(\vec{p},i\omega_p)v_{\vec{p}}^\alpha v_{\vec{p}}^\beta\times\nn\\
&&[\G(\vec{p},\omega + i\eta+i\omega_p)-\G(\vec{p},i\eta+i\omega_p)]. 
\label{sigmaGG}
\earray

We now want to prove that this result holds to each order in $\lambda$. We do this via an inductive argument, in which we assume that through $n^{th}$ order in $\lambda$, $\left(\frac{\delta}{\delta\kappa_i^\alpha}L_x\right)^{(n)}_{\A\to0}$ (where $L_i$ can be $\Psi_i$, $\chi_i$, or $\gamma[i]$) satisfies a certain explicit form (\disp{inducthypkappa}), and then show that this form holds for $n+1^{st}$ order. We then plug \disp{local-ansatzginmukappa} into $\sum_{i-f}\Pi_{\alpha\beta}[i,f]$ (\disp{PiG}), and use the explicit form of $\left(\frac{\delta}{\delta\kappa_i^\alpha}L_x\right)^{(n)}_{\A\to0}$ to simplify the resulting expressions, thereby proving \disp{PiGwGw} and \disp{sigmaGG} to each order in $\lambda$.  

For the reason given below \disp{set1conduct}, we are free to use any of the results from section \ref{EOMlarged}, after making the substitution in \disp{subt}. 
{\bf We define} $X_i$ to be a product of local functions of the type in \disp{objects} (i.e. $X_i=(R_i)^m$) {\bf and} $Y_{i,f}$ to be a {\bf proxy} for either $\GH_{i,f}[\tau_n,\tau_m]$ or $t_{i,\bb{j}}\GH_{\bb{j},f}[\tau_n,\tau_m]$ where the time indices are again arbitrary. {\bf Inductive hypothesis}: Through $n^{th}$ order in $\lambda$, the $\kappa$ source derivative of the objects $\Psi_i$, $\chi_i$, and $\gamma[i]$ (denoted below by the generic symbol $L_i$) can be written as
\barray
 &&\left(\frac{\delta}{\delta\kappa_i^\alpha}L_x\right)^{(n)}_{\A\to0} = \lambda^n \ X_xY_{x,\bb{x_1}}X_{\bb{x_1}}Y_{\bb{x_1},\bb{x_2}}X_{\bb{x_2}}\ldots X_{\bb{x_{m-1}}}\nn\\
 &&Y_{\bb{x_{m-1}},\bb{x_m}}X_{\bb{x_m}}v^{\alpha}_{\bb{x_m},i} \ Y_{i,\bb{x_{m-1}}}X_{\bb{x_{m-1}}}\ldots X_{\bb{x_1}}Y_{\bb{x_1},x}X_x, \nn\\
 \label{inducthypkappa}
\earray
where the number $m$ is arbitrary. In the base case of zeroth order, the objects $\Psi_i$, $\chi_i$, and $\gamma[i]$ are
\barray
&&\Psi^{(0)}_i[\tau_i,\tau_m] = 0; \;\;\;
\gamma^{(0)}[i] = \GH^{(k)}[i,i]; \nn\\
&&\chi^{(0)}_i[\tau_i,\tau_m] = -  (t_{i,\bb{j}}- i\sum_\alpha \kappa_{\bb{j}}^\alpha(\tau_i)v^\alpha_{i\bb{j}})  \ \xi^*. \GH_{\bb{j}, i} [\tau_i, \tau_i] \delta[\tau_i,\tau_m].  \nn\\
\label{basecaseconduct}
\earray
We note that $\left(\frac{\delta}{\delta\kappa_i^\alpha} w_{x,\bb{j}}(\tau_n) \ \GH_{\bb{j}, x} [\tau_n, \tau_m]\right)^{(l)}_{\A\to0}$ is given by \disp{inductstepkappa} with the appropriate objects on the RHS evaluated to the appropriate order in $\lambda$. An analogous formula holds for $\left(\frac{\delta}{\delta\kappa_i^\alpha}\GH_{x,x}[\tau_n,\tau_m]\right)^{(l)}_{\A\to0}$.
Using these formulas with $l=0$ shows that the hypothesis is satisfied for the base case. 

We now prove the inductive step. 
\disp{objects} continues to hold with $t_{i,\bb{j}} \to w_{i,\bb{j}}(\tau_n)$ (the time index is again arbitrary). Therefore, using the notation $\widetilde{R}_i= [R_i]_{t_{i,\bb{j}} \to w_{i,\bb{j}}(\tau_n)}$,  we may write
\beq
\left(\frac{\delta}{\delta\kappa_i^\alpha}L_x\right)^{(n+1)}_{\A\to0} = \sum_{r=0}^{n+1}\lambda^r\left(\frac{\delta}{\delta\kappa_i^\alpha}(\widetilde{R}_x)^m\right)^{(n+1-r)}_{\A\to0}.
\label{psichiderkapnpone}  
\eeq
Substituting the formulas for $\left(\frac{\delta}{\delta\kappa_i^\alpha} w_{x,\bb{j}}(\tau_n) \ \GH_{\bb{j}, x} [\tau_n, \tau_m]\right)^{(l)}_{\A\to0}$ and $\left(\frac{\delta}{\delta\kappa_i^\alpha}\GH_{x,x}[\tau_n,\tau_m]\right)^{(l)}_{\A\to0}$ (\disp{inductstepkappa}) for $l \leq n+1$ into \disp{psichiderkapnpone}, and using the inductive hypothesis, shows that $\left(\frac{\delta}{\delta\kappa_i^\alpha}\Psi_x\right)^{(n+1)}_{\A\to0}$, $\left(\frac{\delta}{\delta\kappa_i^\alpha}\chi_x\right)^{(n+1)}_{\A\to0}$, and $\left(\frac{\delta}{\delta\kappa_i^\alpha}\gamma[x]\right)^{(n+1)}_{\A\to0}$ all have the desired form (\disp{inducthypkappa}). Thus, \disp{inducthypkappa} holds to all orders in $\lambda$. 

Substituting \disp{local-ansatzginmukappa} into $\sum_{i-f}\Pi_{\alpha\beta}[i,f]$ (\disp{PiG}), and using \disp{inducthypkappa}, the only non vanishing terms are those which involve a derivative of the explicit factor $(t_{\bb{x},\bb{y}} - i \sum_\alpha \kappa_{\bb{y}}^\alpha \  v^\alpha_{\bb{x},\bb{y}} )$ from \disp{local-ansatzginmukappa}. The other terms vanish due to the following reasoning. Upon substituting  \disp{inducthypkappa}, in each of these terms there are two paths from $i$ to $f$, both of which pass through the point $\bb{x}$ as well as the points $\bb{x_1}\ldots\bb{x_{m-1}}$ in \disp{inducthypkappa}. Hence, in the large d limit, all of these points must be chosen to be either $i$ or $f$ for these terms to be non vanishing. Then, if we choose $\bb{x_{m-1}}=i$, the term vanishes due to parity, while if we choose $\bb{x_{m-1}}=f$, the term vanishes due to parity combined with the sum $\sum_{i-f}$. Therefore, after making these simplifications, we find that \disp{PiGwGw} and consequently \disp{sigmaGG} hold to each order in $\lambda$.

\subsection{ $O\left(\lambda^2\right)$ theory in the limit of large dimensions\label{integralequations}}
To obtain self-consistent integral equations to any order in $\lambda$ for the objects $\GHI[i,f]$ and $\mu[i,f]$, we expand Eqs. (\ref{local-ansatzginmu}) through (\ref{eq-F}) iteratively in $\lambda$, and set the sources to zero. Once the sources are set to zero, the system becomes translationally invariant in both space and time and we may express the equations in momentum/frequency space. Using the definitions
\beq  \GH_{loc,m}(i\omega_k) \equiv \sum_{\vk} \GH(k)\epsilon_{\vk}^m \label{glocm},\eeq
\barray I_{m_1m_2m_3}(i\omega_k) &\equiv&-\sum_{\omega_p,\omega_q}\GH_{loc,m_1}(i\omega_q)\times\nn\\
&&\GH_{loc,m_2}(i\omega_p)\GH_{loc,m_3}(i\omega_q+ i\omega_p -i\omega_k),\nn\\ 
\earray
the resulting equations to $O\left(\lambda^2\right)$ are:
\beq a_{\G} \equiv 1 - \lambda \frac{n}{2} + \lambda^2\frac{n^2}{4}, \eeq
\barray
 \GHI(k) &=&  i\omega_k + \mbox{\boldmath${\mu'}$} - a_{\G} \ (\varepsilon_k-\frac{u_0}{2})\nn\\
 &&-\lambda\left(\epsilon_{\vk}-\frac{u_0}{2}\right)\Psi(i\omega_k) - \lambda\chi(i\omega_k), \label{ginchi} 
 \earray
\beq \mu(i \omega_k) = a_{\G}+\lambda\Psi(i \omega_k),\label{mumomind} \eeq
\beq \mbox{\boldmath${\mu'}$} = \mbox{\boldmath${\mu}$} -u_0(\lambda\frac{n}{2}-\lambda^2\frac{n^2}{8})+\lambda\sum_p\varepsilon_p\GH(p)- a_{\G} \frac{u_0}{2}, \label{lamexp5id} \eeq
\beq \Psi(i\omega_k) = -\lambda u_0I_{000}(i\omega_k)+2 \lambda I_{010}(i\omega_k),\label{mukidmin} \eeq
\begin{eqnarray} 
\chi(i\omega_k) &=& -\frac{u_0}{2}\Psi(i\omega_k) -u_0 \lambda I_{001}(i\omega_k)\nn\\
&&+2 \lambda I_{011}(i\omega_k). \label{ginkidmin} 
\end{eqnarray}
Before solving the equations, one must set $\lambda =1$. The two Lagrange multipliers \mbox{\boldmath${\mu}$} and $u_0$ are determined by the two sum rules:
\barray
\sum_k\GH(k) =\frac{n}{2}; \;\;\;
\sum_k\G(k)=\frac{n}{2}.
\earray
The objects $\GH_{loc,m}(i\omega_k)$ are given by an appropriate integral over the non-interacting density of states of a function composed of the two self energies $\chi(i\omega_k)$ and $\Psi(i\omega_k)$ and the energy $\epsilon$ (\disp{ginchi}). Therefore, these constitute a self-consistent set of equations for the two self energies. These equations have been solved numerically and compared to DMFT calculations in \refdisp{ECFLDMFT}.

\section{Anderson Model \lab{Anderson}}
A word  is needed at this point on the notation used, since similar looking  symbols represent quite different objects in the \tJ and the AIM models. We use the functions $\G(\{\tau_j \}),\GH(\{\tau_j \}), \mu(\{\tau_j \})$
or $\G(\{ i\omega_j \}), \GH(\{ i\omega_j \}), \mu(\{ i\omega_j \})$ and the related vertex functions for impurity site of the AIM as well, but distinguish them  from the \tJ model variables by dropping the spatial or momentum labels.  Therefore in an equation such as \disp{constraintfourier}, the object on the left (right) hand side corresponds to the \tJ (AIM) models.

\subsection{Equations of Motion for Anderson Model}
In DMFT\cite{GeorgesKotliar,GeorgesKotliaretal}, the local Green's function of the infinite-dimensional finite-U Hubbard model is mapped onto the impurity Green's function of the finite-U AIM, with a self-consistently determined set of parameters. Using the ECFL equations of motion for both models, we show that the same mapping can be made between the infinite-dimensional \tJ model and the infinite-U AIM. Further, we show that this mapping also extends to the auxiliary green's function $\GH$, and the caparison factor $\mu$ individually. In this section, we briefly review the ECFL theory of the AIM\cite{ECFLAM}, and we establish the mapping in the following section. 

Consider the AIM in the limit $U \rightarrow \infty$ which has the following Hamiltonian.
\barray H &=& \sum_\sigma \epsilon_d X^{\sigma\sigma} +\sum_{k\sigma}\widetilde{\epsilon}_kn_{k\sigma} \nn\\
&&+ \sum_{k\sigma}(V_k\ X^{\sigma0} \ c_{k\sigma}+V_k^* \ c^\dagger_{k\sigma} \ X^{0\sigma}),\nn\\
\earray
where we have set the Fermi energy of the conduction electrons to be zero. The impurity Green's function is given by the following expression.
\beq
\G_{\si_i \si_f}[\tau_i,\tau_f]= - \langle\langle \  X^{0 \si_i}(\tau_i) \; X^{\si_f 0}(\tau_f) \label{Green}\rangle\rangle.
\eeq
The ECFL solution of the Anderson model is presented in \refdisp{ECFLAM}. The impurity Green's function is factored into the auxiliary Green's function and the caparison factor.
\beq \G[\tau_i,\tau_f] = \GH[\tau_i,\tau_{\bb{j}}] \ . \mu[\tau_{\bb{j}},\tau_f] \label{GgmuAM}.\eeq
The equations of motion for $\GH$ and $\mu$ can be written as
\barray 
&&(\partial_{\tau_i} +\epsilon_d + \V(\tau_i))\GH[\tau_i,\tau_f] = -\delta(\tau_i-\tau_f)\nn\\
&&-(1-\lambda\gamma[\tau_i]).\Delta[\tau_i,\tau_{\bb{j}}].\GH[\tau_{\bb{j}},\tau_f]\nn\\
&& -\lambda \  \xi^*\Delta[\tau_i,\tau_{\bb{j}}].\GH[\tau_{\bb{j}},\tau_{\bb{x}}].\Lambda_*[\tau_{\bb{x}},\tau_{\bb{y}};\tau_i].\GH[\tau_{\bb{y}},\tau_f], \nn\\
\label{gAM}
\earray
\barray
\mu[\tau_i,\tau_f] &=& \delta(\tau_i-\tau_f)(\iden- \lambda\gamma[\tau_i]) +\nn\\
&& \lambda \ \xi^*.\Delta[\tau_i,\tau_{\bb{j}}].\GH[\tau_{\bb{j}},\tau_{\bb{x}}].\U_*[\tau_{\bb{x}},\tau_f;\tau_i], \label{muAM} 
\earray
where the conduction band enters through the ($\V$ independent) function
\beq \Delta[\tau_i,\tau_f] = -  \iden  \sum_k |V_k|^2 (\partial_{\tau_i}+{\widetilde{\epsilon}}_k)^{-1}\delta(\tau_i-\tau_f).  \eeq
We have also made use of the following definitions:
\barray
&&\Lambda[\tau_n,\tau_m;\tau_i]= - \frac{\delta}{\delta \V(\tau_i)} \GHI[\tau_n,\tau_m];\nn\\
&&\U[\tau_n,\tau_m;\tau_i]=  \frac{\delta}{\delta \V(\tau_i)} \mu[\tau_n,\tau_m];\nn\\
&&\gamma[\tau_i] = \mu^{(k)}[\tau_{\bb{n}},\tau_i^+] . \GH^{(k)}[\tau_i,\tau_{\bb{n}}]. \nn\\
\label{verticesAM} 
\earray
\subsection{Mapping of \tJ model onto Anderson model in infinite dimensions}
Now let us consider the t-J model in the limit of infinite dimensions. Inverting \disp{set1}, the equations of motion for $\GH_{i,i}[\tau_i,\tau_f]$ and $\mu_{i,i}[\tau_i,\tau_f]$ are
\barray 
&&(\partial_{\tau_i} -\mu + \V_i(\tau_i)) \ \GH_{i,i}[\tau_i,\tau_f] =\nn\\
&& -\delta(\tau_i-\tau_f) + (1-\lambda\gamma[i]).\ t_{i,\bb{j}} \ \GH_{\bb{j},i}[\tau_i,\tau_f] +\nn\\
&&\lambda \ t_{i,\bb{j}} \ \xi^*.\GH_{\bb{j},i}[\tau_i,\tau_{\bb{x}}]. A_{i,*}[\tau_\bb{x},\tau_\bb{y};\tau_i]. \GH_{i,i}[\tau_{\bb{y}},\tau_f]\nn \\
&& + \ \lambda \ t_{i,\bb{j}} \ \xi^*.\GH_{\bb{j},i}[\tau_i,\tau_{\bb{x}}]. B_{i,*}[\tau_\bb{x},\tau_\bb{y};\tau_i].\ t_{i,\bb{y}}\ \GH_{\bb{y},i}[\tau_{\bb{y}},\tau_f],\nn\\
\label{gloctJ} 
 \earray
\barray 
\mu_{i,i}[\tau_i,\tau_f] &=& (1- \lambda \gamma[i])\delta(\tau_i-\tau_f) +\nn\\
 && \lambda \ t_{i,\bb{j}} \ \xi^*. \GH_{\bb{j},i}[\tau_i,\tau_{\bb{x}}].B_{i,*}[\tau_\bb{x},\tau_f;\tau_i].\label{muloctJ} 
 \earray
 By mapping $\GH_{i,i}[\tau_i,\tau_f]$ and $\mu_{i,i}[\tau_i,\tau_f]$ onto $\GH[\tau_i,\tau_f]$ and $\mu[\tau_i,\tau_f]$ of the AIM, we would like to show that the equations of motion of the AIM (Eqs. (\ref{gAM}) and (\ref{muAM})) and those of the infinite dimensional \tJ model (Eqs. (\ref{gloctJ}) and (\ref{muloctJ})) map onto each other. To do this, we need the analog of the object $\GHI[\tau_i,\tau_f]$ of the AIM in the \tJ model. We denote this new object $\GH_{loc,i}^{-1}[\tau_i,\tau_f]$ and define it to be the temporal inverse of the local auxiliary Green's function.
\beq \GH_{i,i}[\tau_i,\tau_{\bb{j}}] . \GH_{loc,i}^{-1}[\tau_{\bb{j}},\tau_f]  = \delta(\tau_i-\tau_f). \eeq
Note that  $\GH_{loc,i}^{-1}[\tau_i,\tau_f] \neq \GHI_{i,i}[\tau_i,\tau_f]$. We also define the corresponding vertex.
\beq \Lambda_{loc,i}[\tau_n,\tau_m;\tau_i]= - \frac{\delta}{\delta \V_i(\tau_i)} \GH_{loc,i}^{-1}[\tau_n,\tau_m]. \eeq

We now make use of the following identity.
\barray
 \Lambda_{loc,i}[\tau_x,\tau_{\bb{y}};\tau_i]. \GH_{i,i}[\tau_{\bb{y}},\tau_f] &=& A_{i}[\tau_x,\tau_\bb{y};\tau_i].\GH_{i,i}[\tau_{\bb{y}},\tau_f] +\nn\\ 
&&B_{i}[\tau_x,\tau_\bb{y};\tau_i]. t_{i,\bb{y}} \ \GH_{\bb{y},i}[\tau_{\bb{y}},\tau_f].\nn\\
\label{identity} 
\earray
This identity is easily proven by considering $\frac{\delta}{\delta\V_i(\tau_i)}\GH_{i,i}[\tau_x,\tau_f]$.
\beq \frac{\delta}{\delta\V_i(\tau_i)}\GH_{i,i}[\tau_x,\tau_f]= \GH_{i,i}[\tau_x,\tau_{\bb{j}}]\Lambda_{loc,i}[\tau_{\bb{j}},\tau_{\bb{y}};\tau_i]\GH_{i,i}[\tau_{\bb{y}},\tau_f]. \nn\\ \eeq
The LHS can also be expressed as
\barray
\frac{\delta}{\delta\V_i(\tau_i)}\GH_{i,i}[\tau_x,\tau_f]&=& \GH_{i,i}[\tau_x,\tau_{\bb{j}}] . ( A_{i}[\tau_{\bb{j}},\tau_\bb{y};\tau_i]. \GH_{i,i}[\tau_{\bb{y}},\tau_f] \nn\\
&&+ B_{i}[\tau_{\bb{j}},\tau_\bb{y};\tau_i]. t_{i,\bb{y}} \ \GH_{\bb{y},i}[\tau_{\bb{y}},\tau_f] ).\nn\\
\earray
Left multiplying the above 2 equations by $\GH_{loc,i}^{-1}$, we recover the identity \disp{identity}.
Substituting this identity into \disp{gloctJ}, we obtain
\barray 
&&(\partial_{\tau_i} -\mu + \V_i(\tau_i)) \GH_{i,i}[\tau_i,\tau_f] = \nn\\
&&-\delta(\tau_i-\tau_f)+(1-\lambda\gamma[i]). \ t_{i,\bb{j}}\GH_{\bb{j},i}[\tau_i,\tau_f]\nn\\
&&+\lambda \ t_{i,\bb{j}} \ \xi^*. \GH_{\bb{j},i}[\tau_i,\tau_{\bb{x}}]. \Lambda_{loc,i*}[\tau_{\bb{x}},\tau_{\bb{y}};\tau_i]. \GH_{i,i}[\tau_{\bb{y}},\tau_f]. \nn\\ \label{gloctJ2}
\earray
We are now ready to map the \tJ model onto the Anderson model. To do this, we map the local objects $\GH_{i,i}[\tau_i,\tau_f]$ and $\mu_{i,i}[\tau_i,\tau_f]$ of the \tJ model to the objects $\GH[\tau_i,\tau_f]$ and $\mu[\tau_i,\tau_f]$ of the Anderson model. We also map $\mu$ to $-\epsilon_d$. The following mappings also follow as a consequence of these.
\barray
\gamma[i]&\rightarrow& \gamma[\tau_i]; \;\;\;
 \Lambda_{loc,i}[\tau_n,\tau_m;\tau_i] \rightarrow \Lambda[\tau_n,\tau_m;\tau_i]; \nn\\
&&B_{i}[\tau_n,\tau_m;\tau_i] \rightarrow -\U[\tau_n,\tau_m;\tau_i]. 
 \label{mappings}
\earray
 Comparing \disp{gloctJ2} with \disp{gAM} and \disp{muloctJ} with \disp{muAM}, we see that the equations of motion map onto each other if the following constraint is satisfied.
 \beq t_{i,\bb{j}} \ \GH_{\bb{j},i}[\tau_i,\tau_f] = -\Delta[\tau_i,\tau_{\bb{j}}]. \GH[\tau_{\bb{j}},\tau_f]. \label{self-consist}\eeq
 \subsection{Mapping to each order in $\lambda$}
 The $O(\lambda^2)$ equations for the infinite-dimensional \tJ model and infinite-U AIM are solved numerically in \refdisp{ECFLDMFT} and \refdisp{ECFLAM} respectively. This can in principle be done to higher orders in $\lambda$ as well, and it is therefore interesting to know if the mapping from the previous section holds to each order in $\lambda$. We show that it does, and give a simple prescription for obtaining the ECFL integral equations for one model from those of the other one (\disp{substitutions}). 

 We review the $\lambda$ expansion for the Anderson model from \refdisp{ECFLAM}. There, \disp{gAM} and \disp{muAM} are written as
 \barray
 &&\GHI[\tau_i,\tau_f]= - (\partial_{\tau_i} +\epsilon_d + \V(\tau_i))\delta(\tau_i-\tau_f)\nn\\
 &&-(1-\lambda\gamma[\tau_i]).\Delta[\tau_i,\tau_f]-\lambda\xi^*\Delta[\tau_i,\tau_{\bb{j}}].\GH[\tau_{\bb{j}},\tau_{\bb{x}}].\Lambda_*[\tau_{\bb{x}},\tau_f;\tau_i],\nn\\
\label{ginAM}
 \earray
 \barray
 \mu[\tau_i,\tau_f] &=& \delta(\tau_i-\tau_f)(\iden- \lambda\gamma[\tau_i]) \nn\\ 
 &&+ \lambda\xi^*.\Delta[\tau_i,\tau_{\bb{j}}].\GH[\tau_{\bb{j}},\tau_{\bb{x}}].\U_*[\tau_{\bb{x}},\tau_f;\tau_i]. \label{muAM2} 
 \earray
 The $\lambda$ expansion is obtained in the same way as for the \tJ model, by iterating the equations in $\GHI$ and $\mu$ and keeping track of explicit powers of $\lambda$. The details to $O(\lambda^2)$ can be found in \refdisp{ECFLAM}. To relate this to the $\lambda$ expansion for the infinite-dimensional \tJ model, recall from \disp{objects} that to each order in $\lambda$, $\Psi_i$, $\chi_i$, $\gamma[i]$, $A_i$, and $B_i$ can be written as a product of the functions $\GH_{i,i}[\tau_n,\tau_m]$ and $t_{i,\bb{j}}\GH_{\bb{j},i}[\tau_n,\tau_m]$. We can now state our {\bf inductive hypothesis}: through $n^{th}$ order in $\lambda$, the $\lambda$ expansion for the Anderson model has the form
\barray
\GHI[\tau_i,\tau_m] &=&  -( \partial_{\tau_i} + \epsilon_d + \V(\tau_i)) \ \delta[\tau_i,\tau_m] -  \lambda \ \chi[\tau_i,\tau_m]  \nn\\ 
&&- ( 1- \lambda\gamma[\tau_i])\Delta[\tau_i,\tau_m]  - \lambda \ \Psi[\tau_i,\tau_{\bb{j}}]\Delta[\tau_{\bb{j}},\tau_m],   \nn \\
\mu[\tau_i,\tau_m] &=&  \delta[\tau_i,\tau_m] (1- \lambda\gamma[\tau_i]) +  \ \lambda \Psi[\tau_i,\tau_m],\nn\\
\Lambda[\tau_n,\tau_m;\tau_i] &=& \ A[\tau_n,\tau_m;\tau_i] -  \ B[\tau_n,\tau_{\bb{j}};\tau_i]\Delta[\tau_{\bb{j}},\tau_m],  \nn \\
\U[\tau_n,\tau_m;\tau_i]  &=& - \ B[\tau_n,\tau_m;\tau_i],
\label{local-ansatzAM}
\earray
where through $n^{th}$ order in $\lambda$, the objects $A[\tau_n,\tau_m;\tau_i]$ and $B[\tau_n,\tau_m;\tau_i]$, and through $n-1^{st}$ order in $\lambda$, the objects $\gamma[\tau_i]$, $\chi[\tau_i,\tau_m]$, and $ \Psi[\tau_i,\tau_m]$, can be obtained from their infinite dimensional \tJ model counterparts via the substitution 
 \barray
&&\GH_{i,i}[\tau_n,\tau_m]\rightarrow \GH[\tau_n,\tau_m];\;\;\;
\chem \rightarrow -\epsilon_d;\nn\\
&&t_{i,\bb{j}}\GH_{\bb{j}i}[\tau_n,\tau_m]\rightarrow -\Delta[\tau_n,\tau_{\bb{j}}]. \GH[\tau_{\bb{j}},\tau_m].
 \label{substitutions}
\earray
We first examine the base case of zeroth order.
\barray
A^{(0)}[\tau_n,\tau_m;\tau_i]=  \delta[\tau_i,\tau_n] \delta[\tau_i,\tau_m];\;
B^{(0)}[\tau_n,\tau_m;\tau_i]= 0. \nn\\
\label{basecaseAM}
\earray
Comparing with \disp{basecase}, the hypothesis clearly holds. We now prove the inductive step.
\disp{verticesAM} together with \disp{ginAM} through \disp{local-ansatzAM} imply the following:
\barray
\chi^{(n)}[\tau_n,\tau_m] &=& \xi^*\Delta[\tau_n,\tau_{\bb{j}}].\GH[\tau_{\bb{j}},\tau_{\bb{x}}].A^{(n)}_*[\tau_{\bb{x}},\tau_m;\tau_n], \nn\\
\Psi^{(n)}[\tau_n,\tau_m] &=&- \xi^*\Delta[\tau_n,\tau_{\bb{j}}].\GH[\tau_{\bb{j}},\tau_{\bb{x}}].B^{(n)}_*[\tau_{\bb{x}},\tau_m;\tau_n],\nn\\
A^{(n+1)}[\tau_n,\tau_m;\tau_i]&=& \ \lambda \left(\frac{\delta}{\delta \V(\tau_i)} \chi[\tau_n,\tau_m]\right)^{(n)}, \nn \\
B^{(n+1)}[\tau_n,\tau_m;\tau_i]&=&\lambda \ \delta[\tau_n,\tau_m] \left(\frac{\delta}{\delta \V(\tau_i)} \gamma[\tau_n]\right)^{(n)} \nn\\
&& -\ \lambda \left(\frac{\delta}{\delta \V(\tau_i)} \Psi[\tau_n,\tau_m]\right)^{(n)},\nn\\
\gamma^{(n)}[\tau_i] &=& - \lambda \ \gamma^{(k)(n-1)}[\tau_i]\GH^{(k)}[\tau_i,\tau_i] \nn\\
&&+  \lambda \ \Psi^{(k)(n-1)}[\tau_{\bb{j}},\tau_i]\GH^{(k)}[\tau_i,\tau_{\bb{j}}]. 
\label{AMobjects}
\earray
Comparing with \disp{eq-Clambda}, we see that $\chi^{(n)}[\tau_n,\tau_m] $, $\Psi^{(n)}[\tau_n,\tau_m]$, and $\gamma^{(n)}[\tau_i]$ have the desired form. We also note that
\barray 
\left(\frac{\delta}{\delta \V(\tau_r)} \GH[\tau_i,\tau_m]\right)^{(l)} &=&  \GH[\tau_i,\tau_{\bb{x}}] (A^{(l)}[\tau_{\bb{x}},\tau_{\bb{y}};\tau_r] \nn\\
&&- B^{(l)}[\tau_{\bb{x}},\tau_{\bb{j}};\tau_r]\Delta[\tau_{\bb{j}},\tau_{\bb{y}}] ) \GH[\tau_{\bb{y}},\tau_m].\nn\\
\label{inductstepAM}
\earray
Comparing this with \disp{inductstep} and \disp{inductstep2}, we see that by the inductive  hypothesis, the mapping \disp{substitutions} continues to hold through order $l \leq n$ even after both sides have been acted on with a functional derivative. Furthermore, in evaluating $A^{(n+1)}[\tau_n,\tau_m;\tau_i]$ and  $B^{(n+1)}[\tau_n,\tau_m;\tau_i]$ using \disp{AMobjects}, we will at most need to set $l=n$ in \disp{inductstepAM}. Finally, comparing \disp{AMobjects} with \disp{eq-Alam}, we see that $A^{(n+1)}[\tau_n,\tau_m;\tau_i]$ and  $B^{(n+1)}[\tau_n,\tau_m;\tau_i]$ have the desired form. Thus, we have proven our inductive hypothesis.

Setting the sources to zero, and Fourier transforming \disp{local-ansatzAM}, we may write ($\lambda \to 1$, $\gamma[\tau_i] \to \frac{n_d}{2} \equiv \frac{n}{2}$)
\barray
\GHI(i\omega_k) &=& i\omega_k - \epsilon_d -(1-\frac{n}{2})\Delta(i\omega_k) \nn\\
&&-\chi(i\omega_k) - \Delta(i\omega_k)\Psi(i\omega_k), \nn\\
\mu(i\omega_k) &=& 1 - \frac{n}{2} + \Psi(i\omega_k).
\earray
Comparing with \disp{momindself}, it immediately follows that under the mapping \disp{substitutions}, $\mu_{i,i}(i\omega_k)\to \mu(i\omega_k)$. Furthermore, multiplying both sides of the equation for $\GHI(k)$ by $\GH(k)$, summing over $\vec{k}$, and using the mapping \disp{substitutions}, it follows that $\GH_{i,i}(i\omega_k)\to \GH(i\omega_k)$. Therefore, the ECFL solution of the infinite dimensional \tJ model maps onto the ECFL solution of the AIM to each order in $\lambda$ as long as the following self-consistency condition is satisfied.
\beq \sum_{\vk} \epsilon_{\vk}\GH(k) = \sum_{\vk}\frac{|V_{\vk}|^2}{i\omega_n - \widetilde{\epsilon_{\vk}}}\GH(i\omega_k). \label{constraintfourier} \eeq
This mapping and self-consistency condition can be understood by referring back to DMFT. In DMFT\cite{GeorgesKotliaretal}, the physical Green's function $\G_{i,f}(i\omega_k)$ is determined for any separation of $i$ and $f$ by the local green's function $G_{i,i}(i\omega_k)$ or equivalently the local self energy $\Sigma(i\omega_k)$. The impurity Green's function of the Anderson model $\G(i\omega_k)$ can be set equal to $G_{i,i}(i\omega_k)$ as long as $\widetilde{\varepsilon_k}$ and $V_k$ satisfy a self-consistency condition relating them to $\G(i\omega_k)$ (See Eqs.(13) and (15) of \refdisp{GeorgesKotliaretal}). In the ECFL mapping, the auxiliary Green's function $\GH_{i,f}(i\omega_k)$ is determined for any separation of $i$ and $f$ by the local auxiliary green's function $\GH_{i,i}(i\omega_k)$ and by the local caparison factor $\mu_{i,i}(i\omega_k)$, or equivalently by the two local self energies $\Psi(i\omega_k)$ and $\chi(i\omega_k)$. $\mu_{i,f}(i\omega_k)$ is itself local and related simply to $\Psi(i\omega_k)$. The impurity auxiliary Green's function of the Anderson model $\GH(i\omega_k)$ can be set equal to $\GH_{i,i}(i\omega_k)$ and the caparison factor of the Anderson model $\mu(i\omega_k)$ set equal to $\mu_{i,i}(i\omega_k)$  as long as $\widetilde{\varepsilon_k}$ and $V_k$ satisfy the self-consistency condition \disp{constraintfourier}. We now show that \disp{constraintfourier} can be put into the form of Eqs. (13) and (15) of \refdisp{GeorgesKotliaretal}.  Using \disp{momindself} the LHS can be written as
 \barray \sum_{\vk} \epsilon_{\vk}\GH(k) &=& \frac{-1}{1-\frac{n}{2}+\Psi(i\omega_k)} \times \nn\\
 && [1-(i\omega_k+\chem-\chi(i\omega_k))\GH(i\omega_k)]. \earray
 Using Eqs. (\ref{GrD}), (\ref{GrDys}), (\ref{momindself}), (\ref{selfrelatemin}) and the relation $\G(i\omega_k)= \GH(i\omega_k) . \mu(i\omega_k)$, the above equation becomes
 \beq \Sigma_D(i\omega_k)+\frac{1}{\G(i\omega_k)}-(i\omega_k+\chem) = - \sum_{\vk} \epsilon_{\vk}\GH(k)\frac{1}{\GH(i\omega_k)}. \nn\\ \eeq
 Substituting  \disp{constraintfourier} into the RHS of the above equation, we recover  Eqs.(13) and (15) from \refdisp{GeorgesKotliaretal}.

\section{Conclusion}
In this work we provide a detailed analysis of the simplifications arising from the large dimensionality limit of the  \tJ model, and have given the first few terms in 
the $\lambda$  series that leads to practically usable results.  It is clear  that the formal result of a local Dysonian self energy is already implied by the large $d$ results for  the Hubbard model reviewed in \refdisp{GeorgesKotliaretal},  if we take the limit of infinite $U$;  that is indeed another description of the  model studied here. However it must be kept in mind that the present calculation starts with the infinite $U$ limit already taken,  and thus provides a non trivial check on the uniqueness of the limit of $U \to \infty$ and $d \to \infty$, i.e. its independence on the order  of these two limits.
Also the present work  uses the novel ECFL  methodology that rests on a different set of tools from the ones usually used to study the Hubbard model and its large dimensional limit.  We use the Schwinger equations of motion, as opposed to the usual Feynman-Wick theory, and  we  have obtained analytical results that do not rely on  the Wick's theorem.

Summarizing, we have considered the ECFL theory for the \tJ model ($J=0$)  by establishing the simplifications that arise in the equations of motion  in the limit of large dimensions. The auxiliary Green's function $\GH(k)$ and the caparison factor $\mu(k)$ can be written in terms of two local self energies $\Psi(i\omega_k)$ and $\chi(i\omega_k)$ as in \disp{momindself}. This insight into the structural form of the physical Green's function $\G(k)$ has  been used in a concurrent publication \refdisp{ECFLDMFT},  to benchmark and compare the ECFL and  DMFT calculations. The ECFL integral equations in the large d limit,  derived here  to $O(\lambda^2)$,  have been solved numerically in \refdisp{ECFLDMFT}, and their solution compares favorably with DMFT results. It can be seen explicitly from these equations that \disp{momindself} holds to second order in $\lambda$, with $\Psi(i\omega_k)$ and $\chi(i\omega_k)$ written as a product of the functions $\GH_{loc,m}(i\omega_k)$ (\disp{glocm}) with $m=0$ or $m=1$. This continues to hold to each order in $\lambda$.  We have analyzed the optical conductivity and have shown that it is given by \disp{sigmaGG} in general and to each order in $\lambda$. We have separately  also studied  the ECFL theory of the infinite-U AIM\cite{ECFLAM}, and have shown that there is a mapping between the ECFL of the infinite dimensional \tJ model and the ECFL of the AIM with a self-consistently determined set of parameters (\disp{constraintfourier}). This mapping holds to each order in $\lambda$ and there is a simple prescription for obtaining the ECFL integral equations for one model from those of the other (\disp{substitutions}).

In conclusion,  this work  provides a solid foundation for the study of the \tJ model, and in particular for the ECFL formalism, in the limit of infinite dimensions, by providing exact statements about the k dependence of the self energies, the absence of vertex corrections in computing the conductivity, and finally in yielding  a systematic expansion in the parameter $\lambda$ that enables a quantitative comparison with other methods as in \refdisp{ECFLDMFT}.

\section{Acknowledgements}
We acknowledge stimulating discussions with A. Georges, H. Krishnamurthy and D. Vollhardt.
We also thank  A. Garg and J. Otsuki for helpful comments. This work was supported by DOE under grant no. FG02-06ER46319.

\end{document}